\documentclass[longbibliography,reprint,superscriptaddress,floatfix]{revtex4-2}

\usepackage{graphicx}
\usepackage{mathtools}
\usepackage{amsmath}
\usepackage{textgreek}
\usepackage{txfonts}
\usepackage{xcolor}
\usepackage[colorinlistoftodos,textwidth=3.5cm]{todonotes}
\usepackage[caption=false]{subfig}
\usepackage{color,soul}
\usepackage{tablefootnote}

\begin{document}

\title{Fine structure splitting cancellation in highly asymmetric InAs/InP droplet epitaxy quantum dots}

\author{N. R. S. van Venrooij}
\affiliation{Department of Physics and Astronomy, University of Iowa, Iowa City, 52242, USA}
\affiliation{Department of Applied Physics, Eindhoven University of Technology, Eindhoven 5612 AZ, The Netherlands}

\author{A. R. da Cruz}
\affiliation{Department of Physics and Astronomy, University of Iowa, Iowa City, 52242, USA}
\affiliation{Department of Applied Physics, Eindhoven University of Technology, Eindhoven 5612 AZ, The Netherlands}

\author{R. S. R. Gajjella}
\affiliation{Department of Applied Physics, Eindhoven University of Technology, Eindhoven 5612 AZ, The Netherlands}

\author{P. M. Koenraad}
\affiliation{Department of Applied Physics, Eindhoven University of Technology, Eindhoven 5612 AZ, The Netherlands}

\author{Craig E. Pryor}
\affiliation{Department of Physics and Astronomy, University of Iowa, Iowa City, 52242, USA}

\author{Michael E. Flatt\'e}
\affiliation{Department of Physics and Astronomy, University of Iowa, Iowa City, 52242, USA}
\affiliation{Department of Applied Physics, Eindhoven University of Technology, Eindhoven 5612 AZ, The Netherlands}

\begin{abstract}
We find the single exciton's fine structure splitting (FSS), which splits its  degenerate ground state manifold into singlets, nearly vanishes in highly asymmetric quantum dots  due to the cancellation of splitting effects with markedly different origin. The  dots simulated are those that emerge on top of etch pits through the  droplet epitaxy growth process; these  etch pit dots break square ($C_{4v}$) spatial symmetry, which has been previously associated with small FSS. Configuration interaction calculations predict a vanishing FSS at a specific finite etch pit displacement from the center of the dot, for a structure far from square symmetry.   We thus predict that highly asymmetric quantum dots may still display negligible fine structure splitting, providing new avenues for high-fidelity generation of indistinguishable, polarization entangled photon pairs on demand.
\end{abstract}
\date{\today}
\maketitle

Optically-active quantum dots embedded in a solid-state matrix, which enables  gate control ({\it e.g.} via strain tuning), can provide on-demand emission of indistinguishable, entangled polarization photon pairs as well as other elements of quantum technologies\cite{Stevenson2006, Singh2010, Norris1996, Landin1999}. The fidelity of entangled polarization photon pairs emitted from the dot, however, depends on the energetic splitting (so-called fine-structure splitting, or FSS) between  two ``bright'' exciton states\cite{Stevenson2006, Akopian2006, Huber2018}. 
Lowering the dot symmetry through growth kinetics from square ($C_{4v}$) to asymmetric ($C_{2v}$),  combined with the electron-hole exchange interaction\cite{Bimberg2005, Bayer2002}, commonly provides the main source of this splitting \cite{Bimberg2005,Singh2010}.
Stranski-Krastanov (SK) growth\cite{schneider_hoefling_forchel_2012} relies on surface strain to form quantum dots, therefore producing  dots with highly elongated bases and large FSS \cite{Skiba2017}. Some growth techniques, such as droplet epitaxy (DE)\cite{gurioli2019}, regularly produce embedded quantum dots with near $C_{4v}$ symmetry \cite{Skiba2017, Gajjela2021, Gajjela_Maddalena_2022}; many such dots are more symmetric and have smaller FSS than their SK counterparts \cite{Gajjela2021,Skiba2017}. However DE growth  suffers from the formation of etch pits: secondary structures at the base of the  dot \cite{Keizer2010, Gajjela_2022, Gajjela2021} which can break the structural symmetry of the  dot and potentially increase the FSS. The complex interaction of the quantum dot structure, exchange integrals, and the electron and hole wavefunctions forming the exciton produce several competing effects that have precise names within the literature\cite{klenovsky2015, Takagahara2000}, including long-range exchange, short-range exchange, and band mixing terms; few calculations have attempted to include all relevant terms in the FSS calculation or discuss their interplay.

Here we identify an unexpected cancellation between  FSS terms, denoted in the literature as (i) bulk band mixing, (ii) electric dipole, and (iii) short-range exchange, that emerge from the  Hartree-Fock Hamiltonian evaluated to second order in the electron-hole spatial separation.  As a consequence certain highly asymmetric dots exhibit negligible fine structure splitting.
Our theoretical model utilizes an eight-band ${\bf k}\cdot {\bf p}$ envelope function theory to calculate the bound electron and hole  wavefunctions for realistic quantum dot geometries\cite{Pryor1998}. A configuration interaction (CI) calculation built from these states generates the single exciton energies and wavefunctions\cite{Franceschetti1999}. 
We find that simple etch pit structures  break the exciton wavefunction symmetry and usually increase the FSS, however  certain highly asymmetric etch pit positions  may be beneficial and reduce the FSS to near zero.

\begin{figure}[hbtp!]
\includegraphics[width=0.45\textwidth]{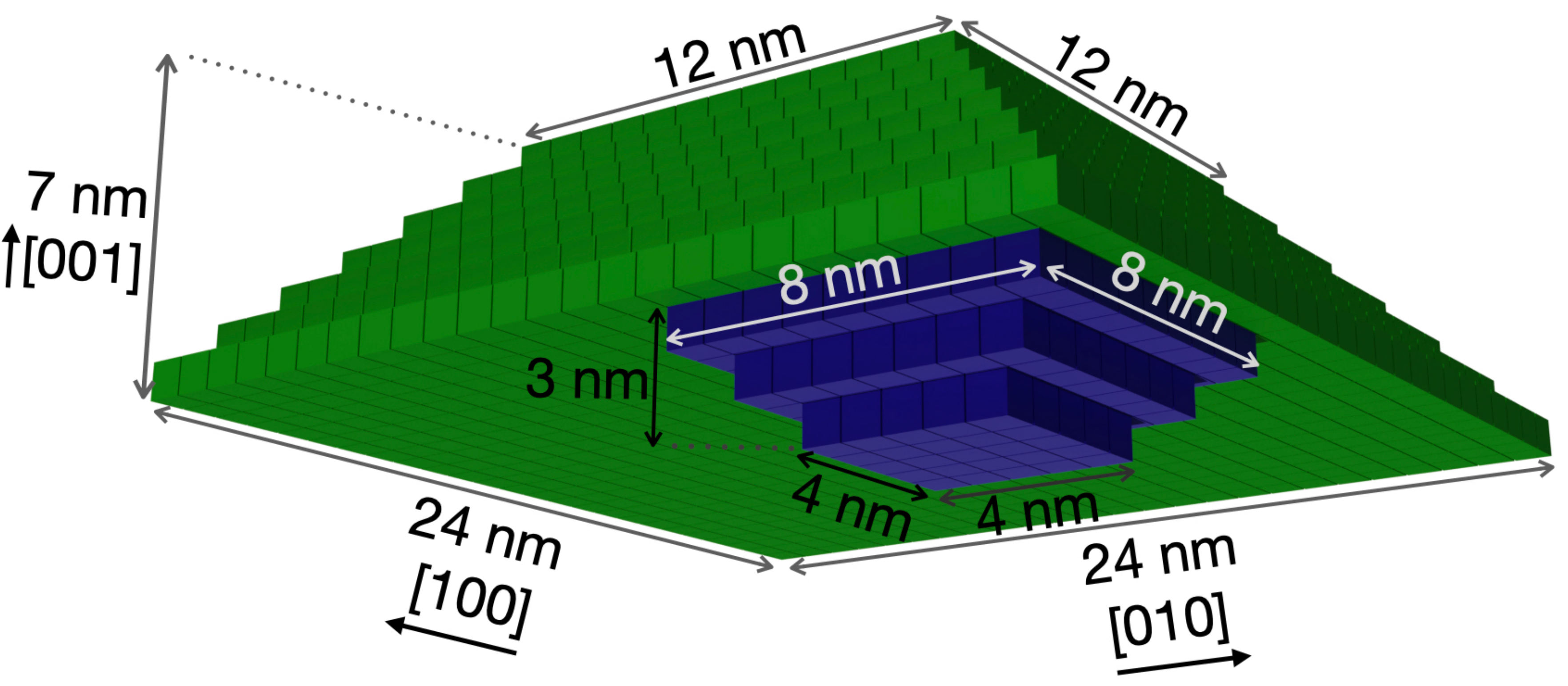}
    \caption{Schematic diagram of the quantum dot (red) and etch pit (blue). Both structures are discretized on a grid with grid spacings of $1\times 1\times 1$ nm\textsuperscript{3}.}
    \label{qd_geom}
\end{figure}

A schematic  of the simulated quantum dot and etch pit is in Fig. \ref{qd_geom}; 
the dot has a base length of 24~nm,  a height of 7~nm, facets on the (101), ($\overline{1}$01), (011) and (0$\overline{1}$1) planes, and a   diagonal baselength  parallel to the (001) plane, coinciding with inferences from scanning tunneling microscopy (STM) measurements\cite{Gajjela_2022}. Because this work is primarily focused on  the effects of etch pits on the excitonic fine structure we consider truncated square base quantum dots and neglect  piezoelectric effects. For such dots all the $C_{4v}$-symmetry-breaking effects are directly related to the positioning of the etch pit. The etch pit shape was  a truncated pyramid\cite{ZOCHER2019219} with a base length of 8 nm and a height of 3 nm. 
The truncation of the pyramid was introduced due to insufficiently definitive data on the etch pit shape,  because the shape of a truncated pyramid is easier to parameterize, and as it roughly approaches the etch pit shape measured experimentally using cross sectional STM \cite{Gajjela_2022}. The quantum dot and etch pit geometries are  projected on a cubic grid with a grid size of 1~nm\textsuperscript{3}.
To study the effect that the  etch pit position has on the quantum dot fine structure the etch pit was shifted along the diagonal to the corner of the quantum dot base in increments of $\sqrt{2}$~nm.

The lowest energy single particle excited states of the dot correspond to adding an electron to the conduction band (``electron'') or removing an electron from the valence band (``hole''). The zone-center Bloch functions of the host material are a complete set of states, hence any conduction or valence band states at finite crystal momentum can be expressed as linear combinations of them. 
The dot's single particle states are calculated using an eight band $k\cdot p$ envelope model described in Ref.~\onlinecite{Pryor1998}. Discrete  states in the quantum dot  are expressed in terms of spatially-varying coefficients (envelope functions) of the Bloch functions associated with the bulk conduction and valence zone-center Bloch functions (below referred to as conduction or valence contributions). Each of those electron and hole states have two spin orientations, and these four states form the smallest basis for a lowest-energy exciton manifold. In the absence of  spin-dependent effects the electron and hole spin degeneracy would produce four degenerate excitons at zero magnetic field. However these four states are split through {electronic} interactions, which are dominated by the exchange interaction. We will thus focus on the four nondegenerate lowest-energy states of a single exciton, including the two spin states of the electron and hole constituents but excluding the biexciton and charged exciton states\cite{Franceschetti1999, Stevenson2006}. 

We orient our discussion of symmetry breaking terms relative to the highest-symmetry case: spherical quantum dots. These have a spin degenerate lowest-energy electron state  ($S=1/2$) and a four-fold degenerate lowest-energy hole state described by total angular momentum (spin and orbit) $J=3/2$ \cite{Sercel1990, Arshak2002}. Without spin-dependent effects this electron and hole degeneracy  produces eight exciton states, corresponding to $J=2$ and $J=1$. The symmetry of DE quantum dots is much lower; 
ideal DE quantum dots are truncated pyramids with a perfectly square base, described by  $C_{4v}$ symmetry, which breaks the four fold degeneracy of the lowest-energy hole. Depending on the quantum dot dimensions either the heavy hole (HH) or the light hole (LH) has lower energy. It is reported that short and wide quantum dots have the HH close to the band gap whereas tall and narrow dots have the LH close to the band gap \cite{Jeannin2017}. This work will solely focus on the latter, which are more commonly grown. 

These ideal DE quantum dots, in the presence of electron-electron interaction, exhibit four nondegenerate  exciton eigenstates composed out of combinations of the conduction band electrons and HH-band dominant holes. Among these excitons there is a (near) degenerate high energy pair with total angular momentum corresponding to $J\approx 1$ and a (near) degenerate low energy pair with total angular momentum corresponding to $J\approx 2$. 
The order occurs because the electron-hole exchange interaction  becomes repulsive as $J \rightarrow 1$ and attractive as $J \rightarrow 2$. These total angular momentum values imply  that all excitons have allowed optical transitions, however the oscillator strengths of the high energy excitons greatly exceed those of the low energy excitons, and thus  the high pair is often denoted as the bright pair whereas the low energy pair is referred to as the dim pair. Self assembled quantum dots will  always have some effective elongation in one of the base diagonals due to strain-induced effects like piezoelectric fields \cite{Bimberg2005}. This lowering of the symmetry further breaks the degeneracy of the high energy excitons and introduces a fine structure splitting. 

The electron and hole wavefunctions in this study are computed using strain-dependent eight band envelope function theory 
on a real space grid. 
The QD electron and hole states can be written as a product between Bloch waves \{$u(\mathbf{r})$\} and spatially-varying envelope functions \{$F(\mathbf{r}), G(\mathbf{r})$\}. The envelopes themselves are approximately constant within one unit cell of the  grid and depend on the discrete grid coordinate $\mathbf{R}$. The Bloch functions  vary over a unit cell and are periodic, so they depend solely on the continuous coordinate $\mathbf{\Tilde{r}}$ within a unit cell. The position vector may then be written as $\mathbf{r} = \mathbf{R}+\mathbf{\Tilde{r}}$. We then compute the confined electron and hole states, which depend on the composition of the quantum dot and its geometry. The electron and hole wavefunctions are

\small
\begin{equation}
\begin{aligned}
\label{eight_band_discrete}
    \psi_e(\mathbf{R},\mathbf{\Tilde{r}}) &= \sum_{i=1}^8 F_i(\mathbf{R}) u_i(\mathbf{\Tilde{r}}),\\
    \psi_h(\mathbf{R},\mathbf{\Tilde{r}}) &= \sum_{j=1}^8 G_j(\mathbf{R}) u_j(\mathbf{\Tilde{r}}),
\end{aligned}
\end{equation}
\normalsize
with $\psi_e$ and $\psi_h$ the electron and hole wavefunctions respectively, $F_i$ the electron envelope functions, $G_j$ the hole envelope functions and $u_i$, the Bloch functions corresponding to each band (using the basis of Ref.~\onlinecite{Bahder1990}, see supplementary material\cite{suppl})\cite{Sakurai1986,Vurgaftman2001,Banin1997,LandoltBornsteinInAs}.
To calculate the eigenenergy of an exciton and to account for the antisymmetrization requirement of a two-body fermionic wavefunction, the wavefunctions in Eq.~(\ref{eight_band_discrete}) are the two-particle Slater determinants \cite{Luo2009,Franceschetti1999}
\small
\begin{equation}
\label{slater_determinant}
\begin{aligned}
    &\Psi_{eh}(\mathbf{R_1},\mathbf{\Tilde{r}}_1,\mathbf{R_2},\mathbf{\Tilde{r}}_2) =\\ &\frac{1}{\sqrt{2}}\Big[\psi_e(\mathbf{R_1},\mathbf{\Tilde{r}}_1)\psi_h(\mathbf{R_2},\mathbf{\Tilde{r}}_2)- \psi_e(\mathbf{R_2},\mathbf{\Tilde{r}}_2)\psi_h(\mathbf{R_1},\mathbf{\Tilde{r}}_1)\Big]=\\
    &\frac{1}{\sqrt{2}}\sum_{i,j=1}^8\Big[F_i(\mathbf{R_1})u_i(\mathbf{\Tilde{r}}_1)G_j(\mathbf{R_2})u_j(\mathbf{\Tilde{r}}_2)-F_i(\mathbf{R_2})u_i(\mathbf{\Tilde{r}}_2)G_j(\mathbf{R_1})u_j(\mathbf{\Tilde{r}}_1)\Big].
\end{aligned}
\end{equation}
\normalsize

An upper bound on the eigenenergy of the exciton is obtained from the expectation value of the Hartree-Fock Hamiltonian,
\small
\begin{equation}
\label{Hartree-Fock}
    \hat{\mathcal{H}}_{HF} = E_e + E_h + \frac{e^2}{4\pi\epsilon_0\varepsilon_\infty}\frac{1}{||\Delta\mathbf{R}+\Delta\mathbf{\Tilde{r}}||},
\end{equation}
\normalsize
with $E_e$ and $E_h$ the eigenenergies of the electron and hole, $e$ the elementary charge, $\epsilon_0$ the vacuum permittivity, $\varepsilon_\infty$ the high frequency (greater than phonon excitation energies) dielectric constant, $\Delta\mathbf{R} = \mathbf{R}_1-\mathbf{R}_2$ and $\Delta\mathbf{\Tilde{r}} = \mathbf{\Tilde{r}}_1-\mathbf{\Tilde{r}}_2$. For a two-particle fermionic wavefunction the Coulomb interaction can be split into the Hartree contribution $J$ and the exchange contribution $K$.

For the four excitons constructed from two electron and two hole states the specific states must be labeled; a specific one particle (electron or hole) or two particle (exciton) state will be labeled with the index $\ell$ to distinguish this label from the band indices $i$ and $j$. The Hamiltonian in Eq.~(\ref{Hartree-Fock}) will mix the four different exciton states  so a matrix Schr\"odinger equation is constructed in the CI calculation. The excitonic eigenfunctions $\Phi$ are written as a linear combination of the Slater determinants in Eq.~(\ref{slater_determinant}):

\small
\begin{equation}
    \label{Exciton_Configuration}
    \Phi_\ell(\mathbf{R}_1,\mathbf{\Tilde{r}}_1,\mathbf{R}_2,\mathbf{\Tilde{r}}_2) = \sum_{\ell'=1}^4 C_{\ell'}\Psi_{\ell'}(\mathbf{R}_1,\mathbf{\Tilde{r}}_1,\mathbf{R}_2,\mathbf{\Tilde{r}}_2).
\end{equation}
\normalsize
 Off diagonal matrix elements between different Slater determinants originate from exchange. The Hamiltonian matrix elements for the CI calculation are 
\small
\begin{equation}
\label{CI_matrix_elements}
\begin{aligned}
    \langle \Psi_{\ell}|\hat{\mathcal{H}}_{HF}|\Psi_{\ell'}\rangle =  &(E_e - E_h)\delta_{\ell,\ell'}\\
    &- J_{\ell,\ell'} + K_{\ell,\ell'},
\end{aligned}
\end{equation}
\normalsize
where 
$J_{\ell,\ell'}$ is the Hartree contribution  and $K_{\ell,\ell'}$ is the exchange contribution. 
Computing these matrix elements and diagonalizing this Hamiltonian results in an upper bound for the exciton eigenenergies.

Within the theoretical framework of an envelope function model the electron-hole exchange interaction interaction is conveniently divided into two main constituents denoted   short range ($SR$), and long range ($LR$) \cite{klenovsky2015, Takagahara2000}, indicating whether the Coulomb interaction occurs between an electron and a hole within the same unit cell or in two different unit cells. The interaction within a unit cell ($\mathbf{R_1}=\mathbf{R_2}=\mathbf{R}$) corresponds to the analytic part of the exchange interaction or the short-range exchange interaction\cite{Bir,Franceschetti1998,Luo2009a,Goupalov1998}. The interaction across different unit cells ($\mathbf{R_1}\neq\mathbf{R_2}$) is referred to as the nonanalytic part of the exchange interaction or the long-range exchange interaction\cite{Franceschetti1998,Luo2009a,Goupalov1998}. A more detailed description of the exchange interaction is given in the supplementary material.

\begin{figure}[tbp!]
    \centering
    \includegraphics[width=0.5\textwidth]{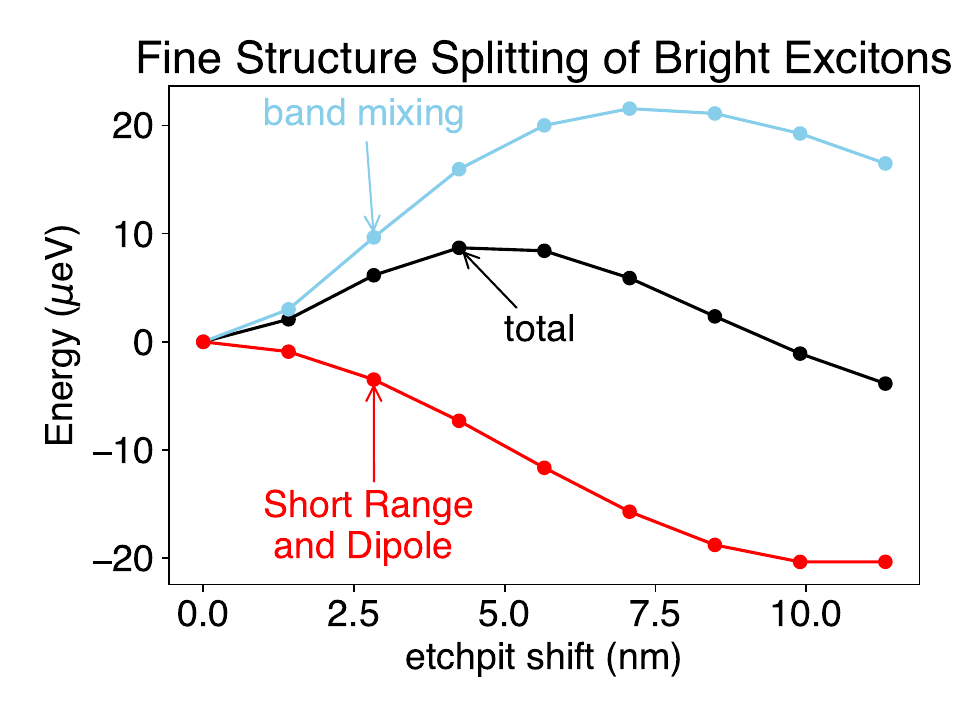}
    \caption{The fine structure splitting between the bright excitons for a droplet epitaxy quantum dot. The fine structure splitting is defined as $FSS_{bright} = E_{brightest}-E_{bright}$. The black line indicates the calculated fine structure splitting for the total exchange interaction. The blue line indicates the calculated fine structure splitting when solely including the band mixing terms of the exchange interaction, these are $K_{LR,BM}^{(0)}$, $K_{LR,BM}^{(1)}$ and $K_{LR,BM}^{(2)}$. The red line indicates the calculated fine structure splitting when solely including the short range ($K_{SR}$) and dipole interactions ($K_{LR,DD}^{(2)}$).}
    \label{FSS}
\end{figure}

Terms in the long range interaction can be further subdivided into  the  second order dipolar interaction ($K_{LR,DD}^{(2)}$) and other, usually neglected, zeroth, first and second order band mixing interactions ($K_{LR,BM}^{(0)}$, $K_{LR,BM}^{(1)}$ and $K_{LR,BM}^{(2)}$ respectively). 
This results into the following expression:

\begin{equation}
\begin{aligned}
	K_{LR} = &K_{LR,BM}^{(0)}(||\Delta\mathbf{R}||^{-1}) + K_{LR,BM}^{(1)}(||\Delta\mathbf{R}||^{-3}) +\\
        &K_{LR,DD}^{(2)}(||\Delta\mathbf{R}||^{-5}) + K_{LR,BM}^{(2)}(||\Delta\mathbf{R}||^{-5}),
\end{aligned}
\end{equation}
where $K_{LR,BM}^{(0)}$, $K_{LR,BM}^{(1)}$ and $K_{LR,BM}^{(2)}$ are respectively  the zeroth, first and second order band mixing interactions. These interactions exclusively occur in models where the electron and hole wavefunctions have  
contributions from both conduction and valence bands. The term $K_{DD}^{(2)}$ is the more frequently used dipole interaction for multiband wavefunctions. Even though the zeroth and first order band mixing terms drop off more slowly than the dipole interaction, the contribution of the dipole interaction to the fine structure splitting is still larger. Nevertheless the band mixing contributions to wavefunctions can be large, even in spherical dots, approaching nearly 30\%\cite{Sercel1990,Bree2014}. 
Thus the band mixing components cannot be ignored  as will become evident below.

The FSS of quantum dots has been previously investigated using the short-range exchange interaction\cite{Efros1996,Landin1999} assuming the electron and hole wavefunctions are composed purely of conduction and valence band components respectively.
In this case the exchange interaction takes the form\cite{Bir}
\small
\begin{equation}\label{SRexch}
\hat H_{exch} = \Delta ~ { \mathbf S }_{\Gamma_6} \cdot  { {\mathbf  J}_{\Gamma_8}} 
\end{equation}
\normalsize
where ${ \mathbf S}_{\Gamma_6}$ is the spin operator for the bottom of the conduction band, $ {\mathbf J}_{\Gamma_8}$ is the spin-$3/2$ operator for the top of the valence band, and the magnitude $\Delta$ is fit to experiment.
This form may be deduced from the fact that $\mathbf S \cdot \mathbf J$ is the only available rotationally invariant operator.

In an eight-band model, however, there are three rotationally invariant operators: $\mathbf S_{\Gamma_6}\cdot \mathbf J_{\Gamma_8}$, $\mathbf S_{\Gamma_6}\cdot \mathbf S_{\Gamma_7}$, and $\mathbf S_{\Gamma_7}\cdot \mathbf J_{\Gamma_8}$, where $\mathbf S_{\Gamma_7}$ is the spin operator for the spin-orbit band.
Each of these three interactions has its own coefficient determined by the integral in {Eq.~}(8) of the supplemental material, which would be fit to experiment.
Lacking sufficient data to do such a fit  forces us to adopt the single band approach conventionally used.

The effect of the etch pit position on the fine structure of the bright excitons can be seen in Fig.~\ref{FSS} which presents the main results of this study. The figure shows the calculated fine structure splitting between the bright exciton states for three different situations; the total exchange interaction, only band mixing interactions and only dipole interactions. 
As the etch pit is shifted further along the diagonal the fine structure splitting gradually increases until it reaches a maximum, then decreases, passes through zero, and turns negative.

\begin{figure}[tbhp!]
    \includegraphics[width=0.5\textwidth]{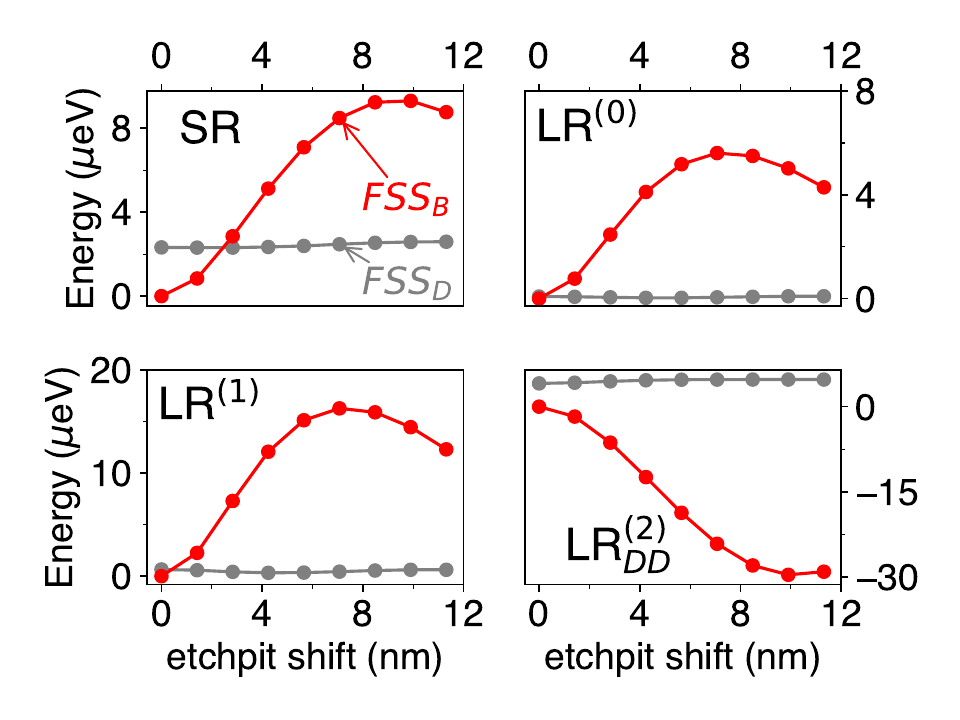}
    \caption{The contributions of each part of the exchange interaction to the fine structure splitting of both the bright (red) and the dark (grey) excitons. It is evident from all plots that the bright states are predominantly affected by the exchange interaction whereas the dark states are not.}
    \label{all_K}
\end{figure}

The origin for this behaviour may be explained by comparing the band mixing contributions to the FSS with the dipole contributions. 
Even though the contributions of both the band mixing terms and the dipole term to the fine structure splitting increase as the etch pit moves from a central location, the FSS induced by these are opposite in sign, causing them to cancel and produce a much smaller total fine structure splitting. 
The attractive sign of the dipole interaction originates from  the very flat geometry of the quantum dot with a large base. 
A more detailed description on how each contribution to the exchange interaction acts separately on the fine structure splitting can be seen in Fig.~\ref{all_K}, however a full discussion of all the nuances is beyond the scope of this work.   Figure~\ref{all_K} identifies the dipole interaction as the largest contributor to the bright exciton FSS. The magnitude of the zeroth order band mixing interaction  is approximately the same as that of the short range exchange interaction, and both are smaller than the magnitude of  the first order band mixing interaction. Thus ignoring the band mixing contributions leads to significantly incorrect exciton fine structure splitting values. 
Figure \ref{all_K} shows that the dark exciton fine structure splitting ($FSS_{dark} = E_{dim} - E_{dark}$) is not  affected nearly as much by the presence of an etch pit as the bright excition fine structure splitting. 

\begin{figure}[ht!]
    \centering
    \includegraphics[width=0.5\textwidth]{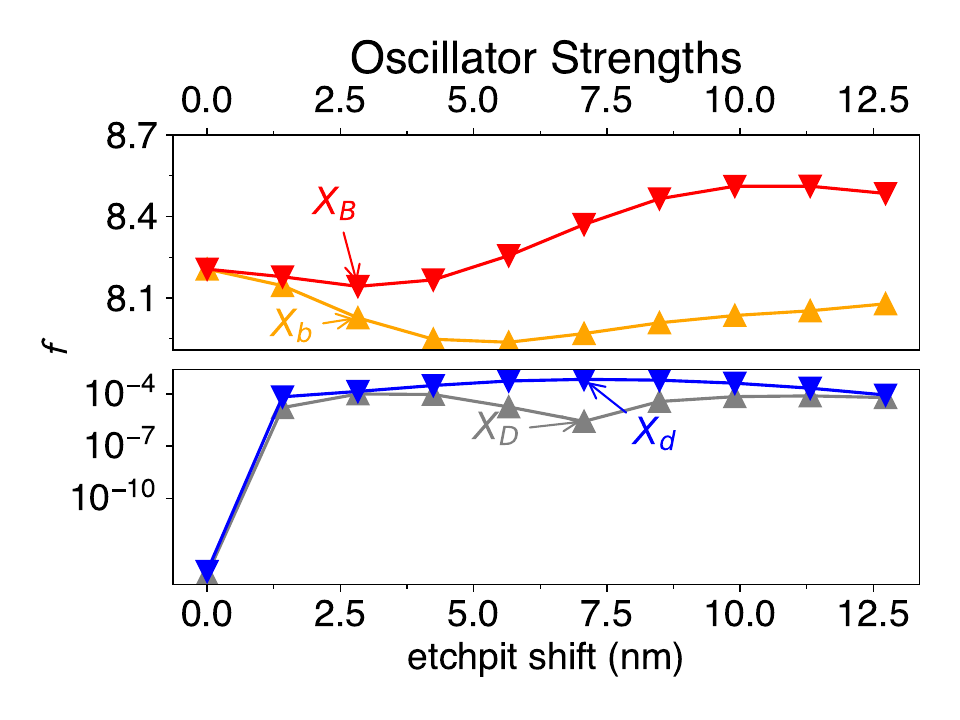}
    \caption{Oscillator strengths for the bright (red) and dark (grey) excitons as a function of the etch pit position. With a centered etchpit the bright excitons have equal oscillator strengths but as the pit shifts one ($X_B$) becomes slightly brighter than the other ($X_b$).  For zero etchpit shift the dark excitons have zero oscillator strength to within the numerical accuracy of the calculations. As the pit shifts, the dark excitons become merely dim ($X_D$ and $X_d$).}
    \label{oscillator}
\end{figure}

The bright versus dark excitons are identified from the oscillator strengths of each exciton, calculated for each etch pit position and shown in Fig.~\ref{oscillator}. The effect of shifting the etch pit  on the exciton oscillator strenghts is dramatic, especially for the dark  states. For a centered etch pit both dark states have a negligible oscillator strength, thus it is not possible to label the states ``dark" and ``dim". As the etch pit is shifted further away from the center, the oscillator strengths increase approximately three orders of magnitude, however remain far smaller than those of the bright states. 
Therefore the dark and bright states do not switch their identification for these dots. By comparing the results of Figs.~\ref{oscillator} and \ref{all_K} we see the presence of an etch pit significantly increases the oscillator strengths of the dark exciton without affecting the dark exciton fine structure splitting. Thus such etch pits may improve dark exciton qubit performance\cite{Holtkemper2021,Zielinski2015}.

\begin{figure}[htbp!]
    \centering
    \includegraphics[width=0.5\textwidth]{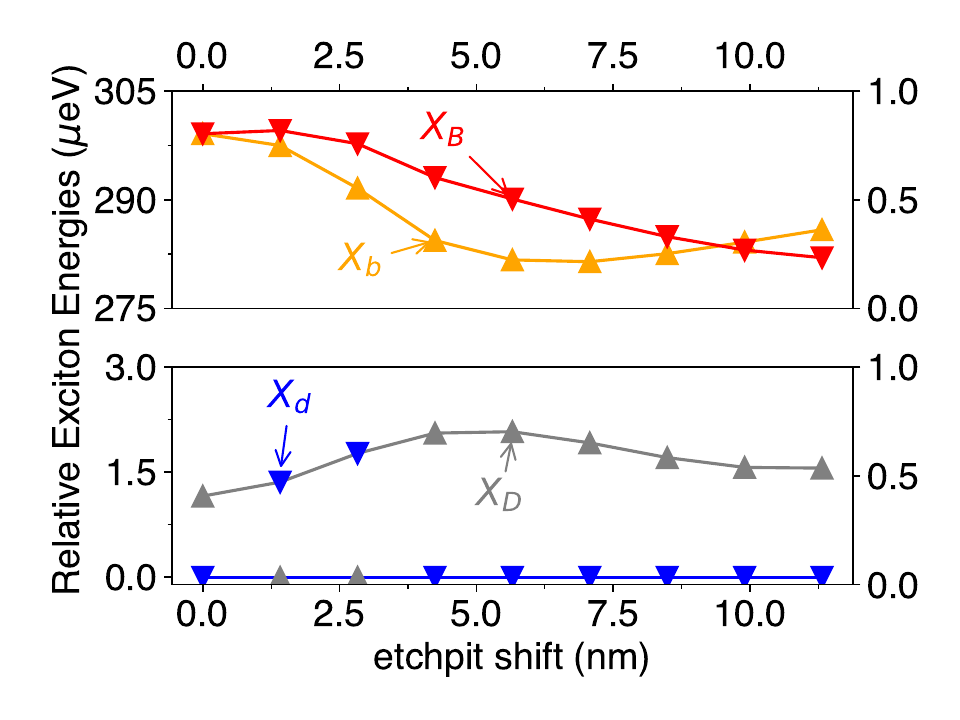}
    \caption{Energies of an exciton in a  quantum dot which has a varying etch pit position, relative to the ground-state exciton. The top figure shows the two bright excitons (${X_b}$, ${X_B}$), which are always higher in energy. The lowest energy exciton can be either the dark exciton ($X_D$) or the dim exciton ($X_d$); The bottom figure shows the splitting between these two, and indicates which is the ground state exciton.}
    \label{Energies}
\end{figure}

\begin{figure}[htbp!]
    \centering
    \includegraphics[width=0.5\textwidth]{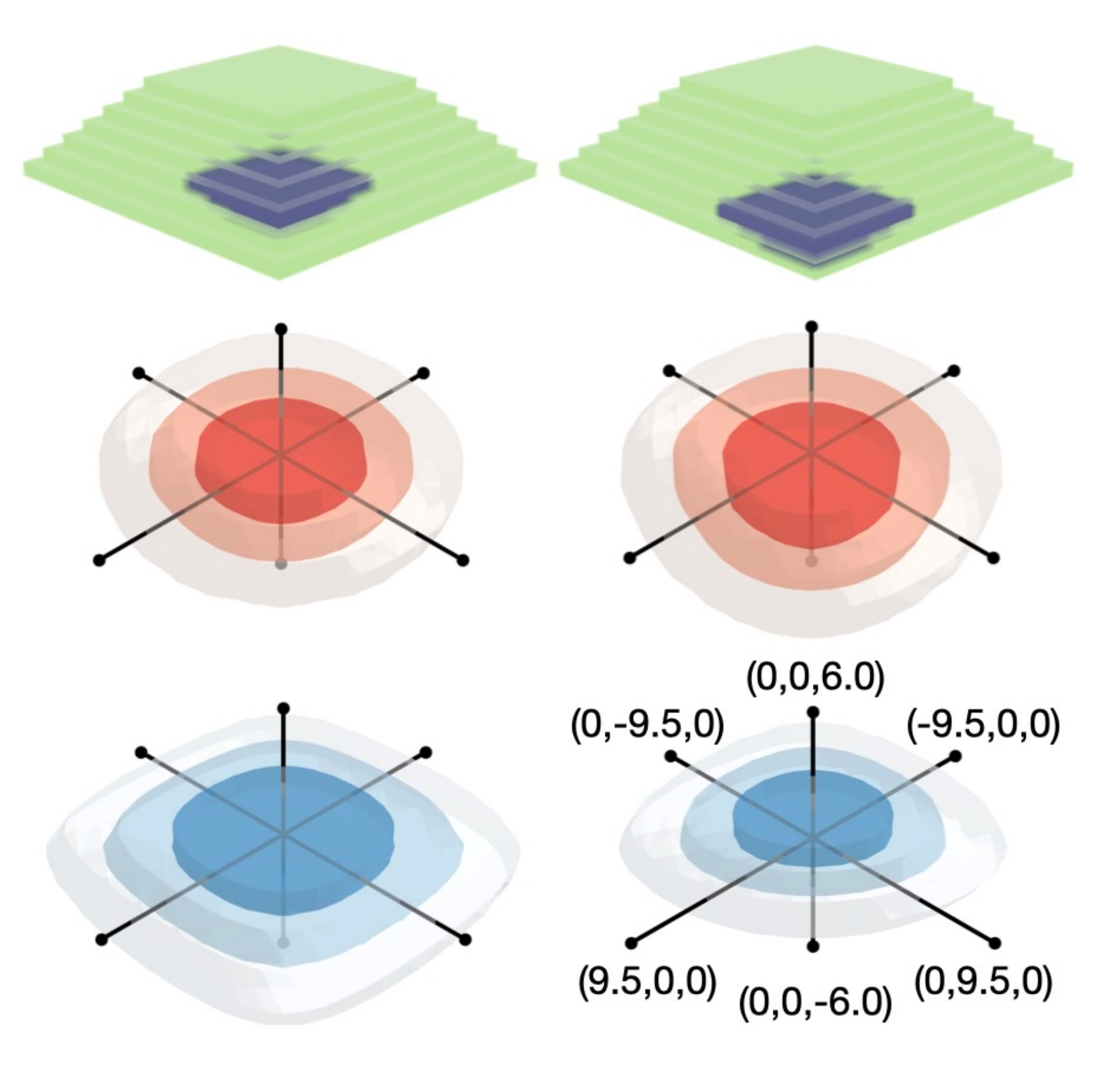} \caption{Single particle electron (red) and hole (blue) states for two different etch pit positions. Above the states the quantum dot and etch pit geometries are depicted schematically (not to scale of the states). The states on the left correspond to a quantum dot with a centered etch pit while the states on the left correspond to a quantum dot with its etch pit shifted along the diagonal. Lines along symmetry axes intersect at the dot center; fixed positions are labeled with black points and their positions relative to the center in units of nm.}
    \label{Qds+states}
\end{figure}

Fig~\ref{Energies} shows the exciton energies with the lowest energy exciton subtracted from the other exciton energies. 
The bright excitons are higher in energy than the dark excitons since the electron-hole exchange interaction acts repulsively on anti-parallel spin configurations.
Fig. \ref{Qds+states} shows maps of the single particle 
states of a quantum dot with an etch pit at two different locations. 
When the etch pit is shifted off center such that $C_{4v}$  symmetry is broken this causes the electron and hole states to respond to this shift by either moving towards the etch pit region (electrons) or away from it (holes).

The degeneracy of the bright excitons is a necessary condition for high fidelity between emitted photons, however the fidelity can be degraded by other effects such as those originating from charge and spin noise. 
As these band mixing terms have led to FSS cancelation in asymmetric structures, we suggest other dot geometries may also produce unexpected features in the exciton fine structure. 

N. R. S. V. and M. E. F. acknowledge useful discussions with M. Atat\"ure and P. Klenovsk\'y. This work was supported by the NSF Grant No. DMR-1921877.

\end{document}